\documentclass[a4papers,11pt]{article}
\usepackage{latexsym,cite,amssymb,amsmath,times}
\usepackage{enumitem}
\usepackage{float}

\input amssym.def
\input amssym.tex

\usepackage{hyperref}

\numberwithin{equation}{section}       % equation numbers in each section

\newcommand{\half}{{{\textstyle\frac{1}{2}}}}

\newcommand{\be}{\setlength\arraycolsep{2pt}\begin{equation}}
\newcommand{\ee}{\end{equation} }
\newcommand{\ba}{\begin{array}}
\newcommand{\ea}{\end{array}}
\newcommand{\bal}{\begin{aligned}}
\newcommand{\eal}{\end{aligned}}
\newcommand{\bpm}{\begin{pmatrix}}
\newcommand{\epm}{\end{pmatrix}}

\newcommand{\comm}[2]{\left[#1,#2\right]}

\newcommand\rd{{\rm d}}

\newcommand\cA{{\cal A}}

\newcommand\cD{{\cal D}}

\newcommand\cF{{\cal F}}

\newcommand\cH{{\cal H}}

\newcommand\cJ{{\cal J}}

\newcommand\cL{{\cal L}}

\newcommand\cP{{\cal P}}

\newcommand\cS{{\cal S}}

\newcommand\bcP{{\bar{\cP}}}

\newcommand\hcL{{\hat{\cal L}}}
\newcommand{\hcLo}{\hat{\cal L}^{\scriptscriptstyle 0}}

\newcommand\hd{\hat{d}}
\newcommand\he{\hat{e}}

\newcommand\hm{\hat{m}}
\newcommand\hn{\hat{n}}
\newcommand\hp{\hat{p}}

\newcommand\hx{\hat{x}}

\newcommand\hB{\hat{B}}
\newcommand\hM{\hat{M}}
\newcommand\hN{\hat{N}}
\newcommand\hP{\hat{P}}
\newcommand\hQ{\hat{Q}}

\newcommand\hT{\hat{T}}

\newcommand\hV{\hat{V}}
\newcommand\hX{\hat{X}}

\newcommand\hmu{\hat{\mu}}
\newcommand\hnu{\hat{\nu}}

\newcommand\hPhi{\hat{\Phi}}

\def\brvare{\bar{\varepsilon}}
\def\breta{\bar{\eta}}

\def\brrho{\bar{\rho}}
\def\brpsi{\bar{\psi}}

\def\brPhi{{{\bar{\Phi}}}}

\def\bra{\bar{a}}
\def\brb{\bar{b}}
\def\brc{\bar{c}}
\def\bre{\bar{e}}

\def\brm{{\bar{m}}}
\def\brn{{\bar{n}}}
\def\brp{{\bar{p}}}
\def\brq{{\bar{q}}}
\def\brr{{\bar{r}}}

\def\brB{\bar{B}}

\def\brP{{\bar{P}}}

\def\brV{{\bar{V}}}

\def\hbe{\hat{\bar{e}}}
\def\hbm{\hat{\bar{m}}}
\def\hbn{\hat{\bar{n}}}
\def\hbp{\hat{\bar{p}}}
\def\hbB{\hat{\bar{B}}}
\def\hbP{\hat{\bar{P}}}
\def\hbV{\hat{\bar{V}}}

\def\hbPhi{\hat{\bar{\Phi}}}

\newcommand{\Spin}{\mathbf{Spin}}

\newcommand{\ODD}{\mathbf{O}(D,D)}
\newcommand{\Odd}{\mathbf{O}(d,d)}
\newcommand{\Onn}{\mathbf{O}(n,n)}

\newcommand{\SpinD}{{\Spin(1,D{-1})}}
\newcommand{\oSpinD}{{{\Spin}(D{-1},1)}}
\newcommand{\Spind}{{\Spin(1,d{-1})}}
\newcommand{\oSpind}{{{\Spin}(d{-1},1)}}
\newcommand{\Spinn}{{\Spin(1,n{-1})}}
\newcommand{\oSpinn}{{{\Spin}(n{-1},1)}}

\newcommand{\Gammao}{\Gamma^{\scriptscriptstyle{0}}{}}

\newcommand{\So}{S^{\scriptscriptstyle{0}}{}}

\begin{document}
\begin{titlepage}
\begin{flushright}
QMUL-PH-13-04
\end{flushright}

\vfill

\begin{center}
   \baselineskip=16pt
   {\LARGE  Supersymmetry for Gauged Double Field Theory \vskip 14pt
    and Generalised Scherk-Schwarz Reductions}
   \vskip 1.5cm
     David S. Berman$^\star$\footnote{\tt d.s.berman@qmul.ac.uk} and Kanghoon Lee $^{\dagger \sharp}$\footnote{\tt kanghoon.lee@imperial.ac.uk}
       \vskip .6cm
             \begin{small}
             	{\it $^\star$Queen Mary University of London, Centre for Research in String Theory, \\
             School of Physics, Mile End Road, London, E1 4NS, England} \\ 
\vspace{2mm}
		{\it $^\dagger$ Department of Physics, Imperial College London \\Prince Consort Road, London, SW7 2AZ, UK }\\
\vspace{2mm}
		{\it $^\sharp$ Center for Quantum Spacetime, Sogang University, Seoul 121-742, Korea}            
\end{small}
\end{center}

\vfill 
\begin{center} 
\textbf{Abstract}
\end{center} 
\begin{quote}
Previous constructions of supersymmetry for double field theory have relied on the so called strong constraint. In this paper, the strong constraint is relaxed and the theory is shown to possess supersymmetry once the generalised Scherk-Schwarz reduction is imposed. The equivalence between the generalised Scherk-Schwarz reduced theory and the gauged double field theory is then examined in detail for the supersymmetric theory. As a biproduct we write the generalised Killing spinor equations for the supersymmetric double field theory.

\end{quote} 
\vfill
\setcounter{footnote}{0}
\end{titlepage}
\newpage
\tableofcontents %%

%%%%%%%%%%%%
\section{Introduction}

Double field theory has been through a recent rebirth. After its orginal inception \cite{Siegel:1993xq,Siegel:1993th} and development \cite{Duff:1989tf,Tseytlin:1990nb,Tseytlin:1990va} there has been a huge number of works by a variety of groups extending the formalism in numerous directions and exploring its consequences \cite{Hull:2004in,Hull:2006qs,Hull:2006va,Hull:2007jy,Hull:2009sg,Hull:2009mi,Hull:2009zb,Hohm:2010jy,
Hohm:2010pp,Hohm:2010xe,Hohm:2011ex,Hohm:2011zr,Hohm:2011dv, Hohm:2011nu,Hohm:2011si, Hohm:2012mf,Jeon:2010rw,Jeon:2011cn,Jeon:2011vx,Jeon:2011sq,Jeon:2011kp,Jeon:2012kd, Jeon:2012hp,Cederwall:2013naa,Cederwall:2013oaa,Berman:2012vc,bermannew,Aldazabal:2013mya,Coimbra:2011nw,Coimbra:2012yy,Coimbra:2012af,Berman:2010is,Berman:2011pe,Berman:2011cg,Berman:2011jh,Berman:2011kg,Dibitetto:2012rk,
Albertsson:2011ux,Copland:2011yh,Thompson:2011uw,Hohm:2011cp,Kan:2011vg,Aldazabal:2011nj,Geissbuhler:2011mx,Grana:2012rr,Berman:2012uy,Musaev:2013rq,West:2010ev,West:2001as,
West:2011mm,Rocen:2010bk,Kwak:2010ew,Copland:2012zz,Kan:2012nf,Geissbuhler:2013uka,Park:2013mpa,Park:2013gaj}. See the following and references therein for a review of the subject \cite{Aldazabal:2013sca,bermanreview}.

In double field theory one doubles the dimension of the space to make the $\ODD$ symmetry manifest on on a 2D dimensional space and then imposes a seperate so called {\it{section condition}} that restricts to a D dimensional submanifold. Different choice of solutions to this section condition produce different T-duality frames. If one may pick a global choice for the solution to the section condition ie. there is a global choice for the T-duality frame then one is ultimately left with a normal supergravity theory and although this reformulation may be interesting we are only rewriting the theory. 

This section condition is intimately tied to the consistency of the theory, the algebra of generalised Lie derivatives depends on the section condition for its closure; the supersymmetric formulations of the theory rely on the section conditions for supersymmetry to work; and various geometric aspects such as tensoral properties appeared to depend directly on the obeying of the section condition.

One of the most exciting aspects of double field theory is to examine to what extent one may relax the section condition and remain a consistent theory. Remarkabley, it is known that the Scherk-Schwarz ansatz allows one to do exactly this \cite{Aldazabal:2011nj,Geissbuhler:2011mx,Grana:2012rr,Berman:2012uy,Musaev:2013rq}. That is we relax the section condition and allow dependence on both the usual coordinaates and their duals simultaneously. However, the geometry is not unconstriained; the generalised metric must obey the so called Scherk-Schwarz factorisation (we will describe this subsequently). It has been shown how all the consistency checks such as closure of the local algebra and the obeying of the Jacobi identity are satisfied even though there is explicit dependence on all the extended coordinates \cite{Aldazabal:2011nj,Geissbuhler:2011mx,Grana:2012rr,Berman:2012uy,Musaev:2013rq}. The generalised Scherk-Schwarz reduced theory then produces a gauged supergravity theory. The embedding tensor \cite{hermann1,hermann2} which determines the gauging then becomes related to the twist matrics of the Scherk-Schwarz anstaz. This result filled a lacuna in M-theory; now all known supergravities theories (with appropriate amounts of supersymmetry) have lifts to a single theory- although that theory neccessarily has novel extended dimensions.

So far there have been different approaches to studying the geometry of these Scherk-Schwarz reduced theories \cite{bermannew,Geissbuhler:2013uka,Aldazabal:2013mya}. In this paper we wish to examine the Scherk-Schwarz reduced theories in the context of the supersymmetric formulation of double field theory developed by \cite{Jeon:2010rw,Jeon:2011cn,Jeon:2011vx,Jeon:2011sq,Jeon:2011kp,Jeon:2012kd, Jeon:2012hp} where one has a semi-covariant formulation (the choices of formalism and their relevant various properties is discussed in \cite{bermannew}).
Using this semicovariant formulation we develop how supersymmetry works in double field theory once we remove the section condition. As a by product we will produce the BPS equations (ie. Killing spinor equations) for double field theory in the absence of section condition. Solving these might have substantial applications for future directions in exploring new and novel solutions to double field theory outside that of usual supergravity.

We begin by describing the geometry for gauged double field theory and then its supersymmetric extension. The generalised Scherk-Schwarz ansatz is described and related to the gauged double field theory in the supersymmetric formalism. Finally we write down the Scherk-Schwarz reduced Killing Spinor equations for double field theory. An extensive appendix gives the details of the bosonic reduction that has appeared elsewhere in the literature (it is repeated here so as to provide notation and a quick reference). In a second appendix the reduced spin connections neccessary for the construction of the reduced Dirac operators are given (this has not appeared before).

In summary and for emphasis, the purpose of this paper is to extend gauged double field theory and the related Scherk-Schwarz reduced double field theory to the supersymmetric case. That is we explicitly construct supersymmetric actions including the Fermionic sector and the associated supervariations.  A key motivation, as stated above, is that in previous supersymmetric constructions of double field theory, the strong constraint appeared as a necessary condition for supersymmetry to work. Here we show that the generalised Scherk-Schwarz case is also consistent with supersymmetry.  This in turn supports the idea that the additional coordinates are physical ie. we can allow nontrivial coordinate dependence in these novel directions (although restricted to be of Scherk-Schwarz type). The details of both the supersymmetric extension of the Scherk-Schwarz reduced theory and the supersymmetrised double gauged supergravity have not appeared previously beyond the bosonic sector. Supersymmetrising double field theory is sufficiently nontrivial that the success of this should not be taken for granted, as such, it is instructive to see how the details work.

%%%%%%%%%%%%
\section{Geometry for Gauged Double Field Theory}

\subsection{Gauged double field theory}
As explained in the introduction, this paper is motivated by seeing how one can remain consistent and yet relax the the physical section condition. In previous work the section condition was crucial for different aspects of the theory to work; this includes the local algebra of generalised diffeomorphisms and importantly supersymmetry. In what follows we will review how imposing the section condition can instead be replaced by the Scherk-Schwarz ansatz. This in turn was then shown to be equivalent to gauging the theory.

And so we start by recalling the gauged double field theory \cite{Hohm:2011ex, Grana:2012rr}. Essentially, it is the gauged double field theory (with its full supersymmetric extension) that we wish to compare with the Scherk-Schwarz reduced double field theory. The reader is encouraged to read \cite{Hohm:2011ex, Grana:2012rr} for the full story. What follows in this section is a brief summary of what appears in those papers so as to define conventions and provide a starting point for introducing the Fermions later.

 Let $V^{M}{}_{N}$ be an arbitrary rank-2 tensor for gauged DFT. $M,N,P$ indices always denote $O(D,D)$ indices with lower case, $m,n,p$ etc. reserved for ordinary $O(d)$ indices.

The gauge symmetry for gauged DFT is given by a twisted generalised Lie derivative which is defined by
\be\ba{rl}
&(\hcL_{X} V)^{M}{}_{N} = (\hcLo_{X} V)^{M}{}_{N} - f^{M}{}_{PQ}X^{P} V^{Q}{}_{N}  - f_{NP}{}^{Q} X^{P}V^{M}{}_{Q}\,,
\\
&\hcL_{X} d = \hcLo_{X} d\, .
\label{symm}
\ea\ee
$\hcLo_{X}$ is the ordinary generalised Lie derivative defined in ungauged DFT by, 
\be\ba{ll}
(\hcLo_{X} V )^{M}{}_{N} = X^{P} \partial_{P} V^{M}{}_{N} + (\partial^{M} X_{P} - \partial_{P} X^{M}) V^{P}{}_{N} + (\partial_{N} X^{P} - \partial^{P} X_{N}) V^{M}{}_{P}\,,
\\
\hcLo_{X} d = X^{M} \partial_{M} d - \half \partial_{M} X^{M}\,,
\ea\ee
where $f_{MNP}$ are the structure constants for Yang-Mills gauge group.  The parameter $X^{M}$ consists of ordinary generalised Lie derivative part and a Yang-Mills gauge symmetry part in an $O(D,D)$ covariant way. The adjoint representation for the gauge parameter $X_{M}{}^{N}$ by may introduced as follows
\be
X_{M}{}^{N} = f_{MP}{}^{N} X^{P}\,, ~~~~~\text{with} ~~~~~ X_{MN} = - X_{NM} \, .
\ee
Then the previous generalised Lie derivatives may be written in the following suggestive form,
\be
(\hcL_{X} V)^{M}{}_{N} = (\hcLo_{X} V)^{M}{}_{N} - X^{M}{}_{P} V^{P}{}_{N}  + V^{M}{}_{P} X^{P}{}_{N} = (\hcLo_{X} V)^{M}{}_{N} + \comm{V}{X}^{M}{}_{N}\,.
\label{symm}
\ee
For consistency of the algebra (ie. closure), arbitrary fields and gauge parameters are required to obey the section condition as in the ordinary DFT. The section condition also known as the strong constraint is given by: 
\be
\partial_{M} \partial^{M} \Phi = 0\,, ~~~~~~~~~~~\partial_{M} \Phi_{1} \, \partial^{M} \Phi_{2} = 0\,
\ee
The structure constants $f_{MNP}$ should then satisfy the Jacobi identity,
\be
f_{M[N}{}^{P} f_{|P|QR]} = 0\,.
\ee 
It is also convenient to impose an orthogonality condition on the structure constants $f_{MNP}$ 
\be
f_{MNP} \, \partial^{M} X = 0\,,
\ee
This means the gauge symmetry will be orthogonal to the ordinary generalised Lie derivative.

Remarkabley one may write the action for gauged double field theory in a very compact form as follows:
\be
\cL_{\rm GDFT} = e^{-2d} \left( \So_{MN} \cH^{MN} + V \right) \, .
\label{gaction}\ee
 $\So_{MN} \cH^{MN}$ is the generalised Ricci scalar for ungauged DFT, and V is the potential for gauged DFT as given in \cite{Geissbuhler:2011mx, Aldazabal:2011nj},
\begin{subequations}
\begin{align}
V_{\rm half-max} =& - \tfrac{1}{2} \cH^{MP} \cH^{NQ} f^{R}{}_{MN} \partial_{P} \cH_{QR} -\tfrac{1}{12} \cH^{MN} \cH^{PQ} \cH^{RS} f_{MPR} f_{NQS} \label{halfmax}
\\&+ \tfrac{1}{4} \cH^{PQ} f_{MNP} f^{MN}{}_{Q} + \tfrac{1}{6} f_{MNP} f^{MNP}\,, \nonumber
\\
V_{\rm max} =&  - \tfrac{1}{2} \cH^{MP} \cH^{NQ} f^{R}{}_{MN} \partial_{P} \cH_{QR} -\tfrac{1}{12} \cH^{MN} \cH^{PQ} \cH^{RS} f_{MPR} f_{NQS} \label{max}
\\&+ \tfrac{1}{4} \cH^{PQ} f_{MNP} f^{MN}{}_{Q} \,.\nonumber 
\end{align}
\end{subequations}
 In the following sections, we will construct the gauged double field theory action (\ref{gaction}) in terms of geometric quantities for gauged double field theory.

%%%%
\subsection{Connection}

To construct a geometry we must make a choice of connection. There are various possibilities depending on what properties one requires of the connection. In \cite{bermannew} a connection is produced that is a full proper connection for the local generalised diffeomorphisms and has the neccessary properties of being $O(D,D)$ compatible and also metric compatible. The price is that it is a flat connection and is torsionful. Unfortunately, that connection does not have nice properties under local $O(D)\times O(D)$ Lorentz transformations (even though the action is indeed invariant as it must be). One of the main motivations of this paper is supersymmetry where the local Lorentz tranformations are crucial. Thus in what follows we will use the so called {\it{semi-covariant}} formulation. This has the price that, as the name suggests, the covariant derivative formed with this connection is not fully covariant under generalised Lie deriviatives. However after a projection it becomes fully convariant and so the semi-covariant derivative in conjuction with a projection operator can be used to construct the fully covariant theory. 

And so, we follow exactly the construction given in \cite{Jeon:2010rw, Jeon:2011cn } for non gauged DFT but now in this paper we will introduce a {\it semi-covariant} derivative for a {\it{twisted generalised Lie derivative}} (\ref{symm}) which will be appropriate for a gauged theory. The {\it{semi-covariant}} derivative acts on a generic quantity carrying $O(D, D)$ vector indices as follows
\be 
  \nabla_{M} T_{N_{1} \cdots N_{n}} 
  = \partial_{M} T_{N_{1} \cdots N_{n}} 
  - \omega \Gamma^{P}{}_{PM} T_{N_{1} \cdots N_{n}}
  + \sum_{m=1}^{n} \Gamma_{M N_{m}}{}^{P} T_{N_{1} \cdots N_{m-1} P N_{m+1} \cdots N_{n}}\, .
\label{covder}
\ee
$\omega$ is a weight factor of each tensor $T_{N_{1}\cdots N_{n}}$  and $\Gamma_{PMN}$ is the connection piece. To determine the connection we assume the following set of constraints exactly analogous with ungauged DFT: 

First, we assume that the semi-covariant derivative preserves the  $\ODD$ metric $\cJ_{MN}$,
\be
\nabla_{M} \cJ_{NP} = \Gamma_{MN}{}^{Q} \cJ_{QP} + \Gamma_{MP}{}^{Q} \cJ_{NQ} = \Gamma_{MNP} + \Gamma_{MPN} = 0\,,
\ee
then it follows that the connection is anti-symmetric for last two indices
\be
\Gamma_{PMN} = \Gamma_{P[MN]}\,.
\ee
Second, we impose the compatibility condition for all NS-NS sector fields,
\be
\nabla_{M} P_{NP} = 0\,, ~~~~~~~~ \nabla_{M} \brP_{NP} = 0\,, ~~~~~~~~ \nabla_{M} d := \partial_{M} d - \half \Gamma^{N}{}_{NM} =  0\,.
\ee
where $P_{MN}$ and $\brP_{MN}$ are projections defined as,
\be
P_{MN} = \half (\cJ_{MN} + \cH_{MN}) \,, ~~~~~~~~~~~~~~ \brP_{MN} = \half (\cJ_{MN} - \cH_{MN})\,,
\ee
satisfying
\be
\ba{lll}
P_{AB}=P_{BA}\,,~~&~~\brP_{AB}=\brP_{BA}\,,~~&~~P_{A}{}^{B}\brP_{B}{}^{C}=0\,,\\
P_{A}{}^{B}P_{B}{}^{C}=P_{A}{}^{C}\,,~~&~~\brP_{A}{}^{B}\brP_{B}{}^{C}=\brP_{A}{}^{C}\,,~~&~~P_{A}{}^{B}+\brP_{A}{}^{B}=\delta_{A}{}^{B}\,.
\ea
\label{symP2}
\ee
Further, we require a generalised torsion free condition:
\be
\hcL_{X}^{\nabla} T^{M} - \hcL^{\partial}_{X} T^{M} = - X^{M}{}_{N} T^{N}  \,.
\label{modtor}
\ee
This is a crucial assumption. In the usual formulation there is no torsion, ie. the righthandside is zero. Now we allow torsion but only of it is of the form given by (\ref{modtor}). The righhandside of (\ref{modtor}) is now a gauge transformation of  $T^{M}$. This means the torsion must also be a gauge tranformation. In ordinary DFT language, it means the geometry is torsion free up to gauge transformations. The difference between $\hcL_{X}^{\nabla} T^{M}$ and $\hcL^{\partial}_{X} T^{M}$ gives
\be
\left(\hcL_{X}^{\nabla} - \hcL^{\partial}_{X} \right) T_{M} = \left( \Gamma_{MNP} + \Gamma_{NPM} + \Gamma_{PMN}\right) X^{P} T^{N}\,,
\label{modtor1}\ee
and from (\ref{modtor}) and (\ref{modtor1}), the modified torsion-free condition implies
\be
\Gamma_{[MNP]} = \tfrac{1}{3} f_{MNP}\,.
\ee
The origin of the contributions to $\Gamma_{[MNP]}$ can be thus be seen from the additional gauge terms appearing in (\ref{symm}) as compared to the terms with no gauging, $\hcLo_{X} V$, which are chosen to give vanishing contribution to the torsion.
One can now construct the connection in terms of $P$, $\brP$, $d$ and the structure constants $f_{MNP}$, which satisfy the compatibility conditions and modified torsion free condition. This is our goal, we have a suitable connection from which we can now contruct the gauged theory. We write this explicitly in terms of the usual connection in the non gauged theory,  $\Gammao_{PMN}$, and new terms:
\be  
\Gamma_{PMN} = \Gammao_{PMN} + \left(\delta_{P}{}^{Q} P_{M}{}^{R} P_{N}{}^{S} + \delta_{P}{}^{Q} \brP_{M}{}^{R} \brP_{N}{}^{S} \right) f_{QRS} -\tfrac{2}{3} \left(\cP + \bar{\cP}\right)_{PMN}{}^{QRS} f_{QRS}\,.
\label{conn}\ee
where $\Gammao_{PMN}$ is the connection for ordinary DFT given in \cite{Jeon:2011cn},
\be
\begin{array}{ll}
\Gammao_{PMN}  = & 2(P\partial_{P} P \brP )_{[MN]} + 2 (\brP_{[M}{}^{Q} \brP_{N]}{}^{R} - P_{[M}{}^{Q} P_{N]}{}^{R} ) \partial_{Q} P_{R P} \\
& - \tfrac{4}{D-1} \left(\brP_{P[M} \brP_{N]}{}^{Q} + P_{P[M} P_{N]}{}^{Q}) \left(\partial_{Q}d + (P\partial^{R} P \brP\right)_{[RQ]}\right)\,,
\end{array}
\label{oldconn}
\ee 
and $\cP$ and  $\bar{\cP}$ are rank-six projection operators
\be
\ba{ll}
\cP_{PMN}{}^{SQR}:=& P_{P}{}^{S}P_{[M}{}^{[Q}P_{N]}{}^{R]}+\tfrac{2}{D-1}P_{P[M}P_{N]}{}^{[Q}P^{R]S}\,,
\\
\bcP_{PMN}{}^{SQR}:=&\brP_{P}{}^{S}\brP_{[M}{}^{[Q}\brP_{N]}{}^{R]}+\frac{2}{D-1}\brP_{P[M}\brP_{N]}{}^{[Q}\brP^{R]S}\,,
\ea\ee
which are symmetric and traceless,
\be
\ba{ll}
{\cP_{CABDEF}=\cP_{DEFCAB}=\cP_{C[AB]D[EF]}\,,}~~&~~{\bcP_{CABDEF}=\bcP_{DEFCAB}=\bcP_{C[AB]D[EF]}\,,} \\
{\cP^{A}{}_{ABDEF}=0\,,~~~~\,P^{AB}\cP_{ABCDEF}=0\,,}~~&~~
{\bcP^{A}{}_{ABDEF}=0\,,~~~~\,\brP^{AB}\bcP_{ABCDEF}=0\,.}
\ea
\label{symP6}
\ee
Here the superscript ‘0’ indicates a quantity defined in the higher dimensional parent DFT.

The connection transforms under the (\ref{symm}) as
\be
\ba{ll}
(\delta_{X} - \hcL_{X}) \Gamma_{PMN} =  -2 \partial_{P} \partial_{[M} X_{N]} + \partial_{P} X_{MN} +  2\left(\cP+\bar{\cP} \right)_{PMN}{}^{QRS} \left( \partial_{Q}\partial_{[R}X_{S]} \right)\,,
\\
(\delta_{X} - \hcL_{X}) \nabla_{P} T_{M} =  2\left(\cP+\bar{\cP} \right)_{PMN}{}^{QRS} \left( \partial_{Q}\partial_{[R}X_{S]} \right)T^{N}\,.
\ea\label{varoldconn}\ee
As in ungauged DFT, the derivative (\ref{covder}) combined with the projections can be used to form generate various covariant quantities such as:
\be
\ba{ll}
&P_{M}{}^{P}{\brP}_{N_{1}}{}^{Q_{1}}{\brP}_{N_{2}}{}^{Q_{2}}\cdots{\brP}_{N_{n}}{}^{Q_{n}}
\nabla_{P}T_{Q_{1}Q_{2}\cdots Q_{n}}\,,
\\
&{\brP}_{M}{}^{P}P_{N_{1}}{}^{Q_{1}}P_{N_{2}}{}^{Q_{2}}\cdots P_{N_{n}}{}^{Q_{n}}
\nabla_{P}T_{Q_{1}Q_{2}\cdots Q_{n}}\, .
%\\
%&P^{MP}{\brP}_{N_{1}}{}^{Q_{1}}{\brP}_{N_{2}}{}^{Q_{2}}\cdots{\brP}_{N_{n}}
%{}^{Q_{n}}\nabla_{M}T_{PQ_{1}Q_{2}\cdots Q_{n}}\,,\\
%&\brP^{MP}{P}_{N_{1}}{}^{Q_{1}}{P}_{N_{2}}{}^{Q_{2}}\cdots{P}_{N_{n}}{}^{Q_{n}} %\nabla_{M}T_{PQ_{1}Q_{2}\cdots Q_{n}}\,,
%\\
%&P^{MP}{\brP}_{N_{1}}{}^{Q_{1}}{\brP}_{N_{2}}{}^{Q_{2}}\cdots{\brP}_{N_{n}}{}^{Q_{n}}
%\nabla_{M} \nabla_{P}T_{Q_{1}Q_{2}\cdots Q_{n}}\,,
%\\
%&{\brP}^{MP}P_{N_{1}}{}^{Q_{1}}P_{N_{2}}{}^{Q_{2}}\cdots P_{N_{n}}{}^{Q_{n}}
%\nabla_{M} \nabla_{P}T_{Q_{1}Q_{2}\cdots Q_{n}}\,.
\ea
\label{covariant}
\ee
This is the whole point of the so called {\it{semi-covariant}} formalism. Some of the quantities are not fully covariant but we can build actions by using the fully covariant projected quantities as building blocks. We can now follow the non gauged case and use the newly contructed semi-convaraint derivative in combination with projections to form the fully covariant theory.

%%%%%%%%%%%%%
\subsection{Spin Connections}

In the previous section we have constructed the relevant connection for $O(D,D)$ tensors in the gauged theory. The spinors though will couple to the local Lorentz group and so we need an appropriate spin connection that will allow us to construct covariant (or {\it{semi-covariant}}) Dirac operators.

Again we will follow \cite{Jeon:2011cn}, but now we will have in mind the gauged extension of the theory. Let us consider a local frame. As in the ungauged DFT, we introduce the double local Lorentz group, $\Spin(1,D-1) \times \Spin(D-1,1)$ and corresponding double-vielbeins, $V_{Mm}$ and $\brV_{M\brm}$. These satisfy the following defining properties\cite{Jeon:2011cn},
\be
\ba{llll}
V_{Ap}V^{A}{}_{q}=\eta_{pq}\,,~~~&~~~~
\brV_{A\brp}\brV^{A}{}_{\brq}=\breta_{\brp\brq}\,,\\
V_{Ap}\brV^{A}{}_{\brq}=0\,,~~~&~~~~V_{Ap}V_{B}{}^{p}+\brV_{A\brp}\brV_{B}{}^{\brp}=\cJ_{AB}\,.
\ea
\label{defV}
\ee
Here unbared indices, $m,n,p,q \cdots$, represent $\Spin(1,D-1)$ vectors and bared indices, $\brm, \brn, \brp, \brq\cdots$, represent $\Spin(D-1,1)$ vectors.
Hence the double-vielbeins   form  a pair of  rank-two  projections~\cite{Jeon:2010rw},
\be
\ba{ll}
P_{AB}:=V_{A}{}^{p}V_{Bp}\,,~~~~&~~~~\brP_{AB}:=\brV_{A}{}^{\brp}\brV_{B\brp}\,,
\ea
\ee
and further meet
\be
\ba{llll}
P_{A}{}^{B}V_{Bp}=V_{Ap}\,,~~&~~\brP_{A}{}^{B}\brV_{B\brp}=\brV_{A\brp}\,,~~&~~\brP_{A}{}^{B}V_{Bp}=0\,,~~&~~P_{A}{}^{B}\brV_{B\brp}=0\,.
\ea
\ee

We define the `master' semi-covariant derivative $\cD_{M}$ acting on any arbitrary $\ODD$, $\Spin(1,D-1)$ and $\Spin(D-1,1)$ representations as follows 
\be
\cD_{M} := \partial_{M}  + \Gamma_{M} + \Phi_{M} + \brPhi_{M}\, . 
\ee
 $\Phi_{M}$ and $\brPhi_{M}$ are spin connections for $\Spin(D-1,1)$ and $\Spin(1,D-1)$ respectively. {\bf Note that the connection $\Gamma_{M}$ and the spin connections $\Phi_{M}$ and $\bar{\Phi}_{M}$ contains Yang-Mills connection part in manifestly $O(D,D)$ covariant manner. Therefore the master derivative $\cD_{M}$ is semi-covariant under the twisted generalised Lie derivative (\ref{symm}) for all representations.}

We then impose the generalised vielbein compatibility condition for these double-vielbeins $V_{Mm}$ and $\brV_{M\brm}$,
\be
\cD_{M} V_{Nm} = 0\,, ~~~~~~~~ \cD_{M} \brV_{N\brm} = 0\,,
\label{compv}
\ee
and for the metric of $\Spin(1,D-1)$ and $\Spin(D-1,1)$, $\eta_{mn}$ and $\breta_{\brm\brn}$ respectively,
\be
\cD_{M} \eta_{mn} =  0\,, ~~~~~~~~~~ \cD_{M} \breta_{\brm\brn} = 0\,.
\ee
From the compatibility of $\eta_{mn}$ and $\breta_{\brm\brn}$, we can deduce that the spin-connections are antisymmetric,
\be
\Phi_{Mmn} = \Phi_{M[mn]}\,, ~~~~~~~~ \brPhi_{M\brm\brn} = \brPhi_{M[\brm\brn]}\,.
\ee
In addition, because of the double-vielbein compatibility condtion (\ref{compv}), the spin-connections may be determined in terms of the double-vielbeins as follows,
\be
\Phi_{Mmn} = V^{N}{}_{m} \nabla_{M} V_{Nn}\,, ~~~~~~~~ \Phi_{Mmn} = V^{N}{}_{m} \nabla_{M} V_{Nn}\,,
\label{spinconn}\ee 
and using (\ref{varoldconn}), these spin-connections are semi-covariant as well,
\be
\ba{ll}
(\delta_{X} - \hcL_{X}) \Phi_{Mmn} = 2\cP_{MNP}{}^{QRS}\partial_{Q}\partial_{[R}X_{S]}V^{N}{}_{m}V^{P}{}_{n}\,,
\\
(\delta_{X} - \hcL_{X})  \brPhi_{M\brm\brn} = 2\bar{\cP}_{MNP}{}^{QRS}\partial_{Q}\partial_{[R}X_{S]}\brV^{N}{}_{\brm}\brV^{P}{}_{\brn}\, .
\ea\ee
Crucially, we can then form fully covariant quantities by contracting the semi-covariant quantities with projection operators or double-vielbeins as shown below:
\be
\ba{llllll}
\brP_{M}{}^{N}\Phi_{Npq}\,,~&~P_{M}{}^{N}\brPhi_{N\brp\brq}\,,~&~\Phi_{M[pq}V^{M}{}_{r]}\,,~&~
\brPhi_{M[\brp\brq}\brV^{M}{}_{\brr]}\,,~&~\Phi_{Mpq}V^{Mp}\,,~&~\brPhi_{M\brp\brq}\brV^{M\brp}\,.
\ea
\label{covPhi}
\ee 
This willl be a reoccuring trick that the formalism uses. One produces fully covariant objects by contracting semicovariant objects with projection operators.

%%%%%
\subsection{Curvature}
Again following \cite{Jeon:2010rw,Jeon:2011cn}, we may construct a rank-4 quantity $R_{PQMN}$ which is generated by the commutator of the semi-covariant derivatives but now for the gauged theory,
\be
\comm{\nabla_{M}}{\nabla_{N}} V_{P} = - \Gamma^{Q}{}_{MN} \nabla_{Q}V_{P} + R_{P}{}^{Q}{}_{MN} V_{Q} \, .
\ee
The curvature, $R_{PQMN}$ is given by
\be
R_{PQMN} = \partial_{M}\Gamma_{NPQ} - \partial_{N}\Gamma_{MPQ} + \Gamma_{MP}{}^{R} \Gamma_{NRQ} - \Gamma_{NP}{}^{R} \Gamma_{MRQ} + 3\Gamma_{[RMN]} \Gamma^{R}{}_{PQ}\,.
\ee 
Note, that unlike ordinary DFT, an additional term is introduced in $R_{PQMN}$. Note it also satisfies the same properties as ungauged DFT, namely that,
\be
R_{MNPQ} = R_{[MN][PQ]} \,, ~~~~~~~~~~~~ P_{M}{}^{R} \brP_{N}{}^{S} R_{RSPQ} = 0\, .
\ee
We can then define the semi-covariant curvature, $S_{MNPQ}$, by
\be
S_{MNPQ} = \half \left(R_{PQMN} + R_{MNPQ}-\Gamma_{RMN} \Gamma^{R}{}_{PQ} \right)\,.
\label{curvature}\ee
Just as for an ordinary Riemann curvature tensor, the semi-covariant curvature satisfies the following symmetry properties on its indices,
\be
S_{[MN][PQ]} = S_{MNPQ}\,,~~~~~~~~~~~~~~S_{MNPQ} = S_{PQMN}\, .
\ee
The Jacobi identity for the structure constants implies the Bianchi identities as well,
\be
S_{M[NPQ]} = 0\,.
\ee
The variation of $S_{MNPQ}$ is given by
\be
(\delta_{X} - \hcL_{X}) S_{PQMN} = 2\nabla_{[M} \left((\cP+ \bar{\cP})_{N]PQ}{}^{RST} \partial_{R}\partial_{[S} X_{T]} \right) + 2\nabla_{[P} \left((\cP+ \bar{\cP})_{Q]MN}{}^{RST} \partial_{R}\partial_{[S} X_{T]} \right)\,.
\label{varS}\ee
Even though $S_{MNPQ}$ is not a fully covariant tensor, we can generate proper scalar  objects by contracting with projection operators as folows
\be
P^{MP} P^{NQ} S_{MNPQ} \,, ~~~~~~~~~~ \brP^{MP} \brP^{NQ} S_{MNPQ}\,.
\ee
Note that these scalars are not equivalent to each other. The scalar curvatures can then be rewritten in terms of $P_{MN}, \brP_{MN}$ and $S_{MNPQ}$. The two possible combination are: \
\be\ba{ll}
P^{MP}P^{NQ} S_{MNPQ} =& P^{MP}P^{NQ} \So_{MNPQ}(\Gammao) + P^{MP} P^{NQ} f^{R}{}_{MN} \Gammao_{RPQ}\
\\&+ \tfrac{1}{6} \left(P_{M}{}^{Q} P_{N}{}^{R} P_{P}{}^{S} + 3 \brP_{M}{}^{Q} P_{N}{}^{R} P_{P}{}^{S}\right) f_{MNP} f_{QRS}\,,
\\
\brP^{MP}\brP^{NQ} S_{MNPQ} = & \brP^{MP}\brP^{NQ} \So_{MNPQ}(\Gammao)  +  \brP^{MP} \brP^{NQ} f^{R}{}_{MN} \Gammao_{RPQ}
\\&+ \tfrac{1}{6} \left(\brP_{M}{}^{Q} \brP_{N}{}^{R} \brP_{P}{}^{S} + 3 P_{M}{}^{Q} \brP_{N}{}^{R} \brP_{P}{}^{S}\right) f_{MNP} f_{QRS}\, .
\ea\ee

These two terms however are related, after some work one can show that
\be
P^{MP}P^{NQ} S_{MNPQ} + \brP^{MP}\brP^{NQ} S_{MNPQ} = \tfrac{1}{6} f_{MNP} f^{MNP}\,.
\label{relation}\ee

Now, we have some choices about how we combine these two terms. In fact, different combinations of these two terms will then produce the actions for the half-maximal and maximal gauged supergravity (\ref{halfmax}) and (\ref{max}). 

First, the half-maximal supersymmetric case, the NS-NS sector Lagrangian is given by 
\be\ba{ll}
\cL_{\rm half-max} &= e^{-2d} \left( 2 P^{MP}P^{NQ} S_{MNPQ}\right)
\\
&=e^{-2d} \Big[\left(P^{MP}P^{NQ} - \brP^{MP}\brP^{NQ}\right) S_{MNPQ} + \tfrac{1}{6} f_{MNP} f^{MNP} \Big] \,,
\ea\label{halfmaxaction}\ee
To see this write it in terms of the generalised metric $\cH_{MN}$, (\ref{halfmaxaction}) then becomes
\be\ba{ll}
\cL_{\rm half-max} = e^{-2d} &\left( 2P^{MP}P^{NQ} \So_{MNPQ} - \tfrac{1}{2} \cH^{MP} \cH^{NQ} f^{R}{}_{MN} \partial_{P} \cH_{QR} \right.
\\ &\left. -\tfrac{1}{12} \cH^{MN} \cH^{PQ} \cH^{RS} f_{MPR} f_{NQS} + \tfrac{1}{4} \cH^{PQ} f_{MNP} f^{MN}{}_{Q} + \tfrac{1}{6} f_{MNP} f^{MNP} \right) \,.
\ea\ee
This is exactly same potential as (\ref{halfmax}).

Second, for the maximal supersymmetric case, the NS-NS sector Lagrangian is given by
\be
\cL_{\rm max} = e^{-2d} \left(P^{MP}P^{NQ} - \brP^{MP}\brP^{NQ} \right) S_{MNPQ} \,,
\label{maxaction}\ee 
which again can be rewritten as
\be\ba{ll}
\cL_{\rm max} = e^{-2d} &\left( 2P^{MP}P^{NQ} \So_{MNPQ} - \tfrac{1}{2} \cH^{MP} \cH^{NQ} f^{R}{}_{MN} \partial_{P} \cH_{QR} \right.
\\ &\left. -\tfrac{1}{12} \cH^{MN} \cH^{PQ} \cH^{RS} f_{MPR} f_{NQS} + \tfrac{1}{4} \cH^{PQ} f_{MNP} f^{MN}{}_{Q} 
\right) \,.
\ea\ee
This is the potential of maximal sugra (\ref{max}).

Thus, we have produced two very simple expressions (\ref{maxaction}) and (\ref{halfmaxaction}) for the action of the bosonic sector in terms of the gauged double field theory curvature, connection and projection operators.

%
%%%%%%%%%%%%%%%%
\section{Supersymmetric Gauged Double Field Theory}
We are now ready to consider the full supersymmetric gauged double field theory with half-maximal supercharges from $10D$ minimal superDFT \cite{Jeon:2011sq,Jeon:2012hp}.
The bosonic sector of the supersymmetric gauged DFT consists of DFT-dilaton, $d$, and double-vielbeins, $V_{Mm}$, $\brV_{M\brm}$.

The fermionic degrees of freedom are given by the gravitino, $\psi^{\alpha}_{\brp}$ and the dilatino, $\rho^{\alpha}$, where $\alpha\,, \beta\,, \cdots$ represent $\Spin(1,9)$ indices. The $\Spin(1,9)$ Clifford algebra,
\be
(\gamma^{m})^{*} = \gamma^{m} \,, ~~~~~~~~ \gamma^{m} \gamma^{n} + \gamma^{n} \gamma^{m} = 2 \eta^{mn}\,,
\ee
and chirality operator $\gamma^{(11)} = \gamma^{0} \gamma^{1} \cdots \gamma^{9}$. 
The symmetric charge conjugation matrice,  $C_{\alpha\beta} = C_{\beta\alpha}$, meets
\be
(C\gamma^{p_{1}p_{2}\cdots p_{n}})_{\alpha\beta}=(-1)^{n(n-1)/2}(C\gamma^{p_{1}p_{2}\cdots p_{n}})_{\beta\alpha}\,,
\ee
and define the charge-conjugated spinors,
\be
\brpsi_{\brp\alpha}=\psi_{\brp}^{\,\beta}C_{\beta\alpha}\,,~~~~~~~~
\brrho_{\alpha}=\rho^{\beta} C_{\beta\alpha}\,.
\ee
The gravitino and dilatino are set to be Majorana-Weyl spinors of the fixed chirality,
\be
\gamma^{(11)} \psi_{\brp} = \psi_{\brp} \,, ~~~~~~~~~~ \gamma^{(11)} \rho = -\rho\,.
\ee
The following table summarises the field content of the half-maximal supersymmetric gauged DFT.
\begin{table}[H]
\begin{center}
\begin{itemize}
\item Bosons 
\begin{itemize}
\item NS-NS sector $~\left\{
\ba{lll}
\mbox{DFT-dilaton:}~~~&~~~\,d\\
\mbox{Double-vielbeins:}~~~&~~~V_{Ap}\,,~~\quad\brV_{A\brp}
\ea
\right.$
\end{itemize}
\item Fermions
\begin{itemize}
\item DFT-dilatino:  $\quad\rho^{\alpha}\,,$
\item Gravitino:~~~~~ $\quad\psi^{\alpha}_{\brp}\,.$
\end{itemize}
\end{itemize}
\label{TABfields}
\end{center}
\caption{Field Contents}
\end{table}

The Dirac operators for $\Spin(1,9)$ spinors are denoted by \cite{Jeon:2011vx}
\be
\gamma^{m} \cD_{m} \rho\,, ~~~~~~~~\cD_{\brm} \rho\,, ~~~~~~~~ \gamma^{m} \cD_{m} \psi_{\brn}\,.
\ee
The explicit form for these is then given by,
\be\ba{rl}
\gamma^{m} \cD_{m} \rho =& \gamma^{m} \partial_{m} \rho + \tfrac{1}{4} \Phi_{mnp}\gamma^{mnp} \rho + \half \Phi^{m}{}_{mp} \gamma^{p} \rho\,,
\\
\cD_{\brm} \rho =& \partial_{\brm} \rho + \tfrac{1}{4} \Phi_{\brm np} \gamma^{np} \rho\,,
\\
\gamma^{m}\cD_{m} \psi_{\brn} = & \gamma^{m} \partial_{m} \psi_{\brn} + \tfrac{1}{4} \Phi_{mnp}\gamma^{mnp} \psi_{\brn} + \half \Phi^{m}{}_{mp} \gamma^{p} \psi_{\brn} + \gamma^{m} \brPhi_{m \brn\brp} \psi^{\brp}\,.
\ea
\ee
Since the Dirac operators use the covariant spin-connections (\ref{covPhi}), these are all invariant under the full gauged DFT symmetries.
We will divide the Dirac operators as ungauged part plus additional terms introduced by gauging, this shows the parts being introdiced by the gauging proceedure in the DFT. 
\begin{equation}
\begin{array}{rl}
       \gamma^{m} \cD_{m} \rho =& \gamma^{m} \cD^{0}_{m} \rho + \frac{1}{12} f_{MNP} V^{M}{}_{m} V^{N}{}_{n} V^{P}{}_{p} \gamma^{mnp} \rho \,,
\\
\cD_{\brm} \rho =& \cD^{0}_{\brm} \rho + \frac{1}{4} f_{MNP}\bar{V}^{M}{}_{\bar{m}} V^{N}{}_{n} V^{P}{}_{p} \gamma^{np} \rho \,,
\\
\gamma^{m}\cD_{m} \psi_{\brn} = & \gamma^{m}\cD^{0}_{m} \psi_{\brn} +\frac{1}{12} f_{MNP}V^{M}{}_{m} V^{N}{}_{n} V^{P}{}_{p} \gamma^{mnp} \psi_{\brn} +  f_{MNP}V^{M}{}_{m} \bar{V}^{N}{}_{\bar{n}} \bar{V}^{P}{}_{\bar{p}} \gamma^{m} \psi^{\bar{p}}\,,

\end{array}
\end{equation}
where $\cD^{0}_{M}$ is the master derivative for ungauged DFT.

We are now in a position to construct a supersymmetric action with half-maximal supersymmetry as follows.
\be
\cL_{\rm \scriptscriptstyle SGDFT} = e^{-2d} \Big[2 P^{MP} P^{NQ} S_{MNPQ} + 4i \left(\brrho \gamma^{m} \cD_{m} \rho -2 \brpsi^{\brm} \cD_{\brm} \rho -   \brpsi^{\brm} \gamma^{m}\cD_{m} \psi_{\brm}\right)\Big]\, .
\label{SGDFT}\ee
Again to demonstrate what is new we can write this action in terms of the ungauged part plus additional terms that come from gauging.
\begin{equation}
\begin{array}{ll}
 \cL_{\rm \scriptscriptstyle SGDFT}  = e^{-2d} \Big[ & 2 P^{MP} P^{NQ} S^{0}_{MNPQ} (\Gamma^{0}) - \tfrac{1}{2} \cH^{MP} \cH^{NQ} f^{R}{}_{MN} \partial_{P} \cH_{QR} 
\\&
-\tfrac{1}{12} \cH^{MN} \cH^{PQ} \cH^{RS} f_{MPR} f_{NQS} + \tfrac{1}{4} \cH^{PQ} f_{MNP} f^{MN}{}_{Q} + \tfrac{1}{6} f_{MNP} f^{MNP} 
\\&
 + 4i \left(\brrho \gamma^{m} \cD^{0}_{m} \rho -2 \brpsi^{\brm} \cD^{0}_{\brm} \rho -   \brpsi^{\brm} \gamma^{m}\cD^{0}_{m} \psi_{\brm} \right) 
\\&
+ i  \frac{1}{3} f_{MNP} V^{M}{}_{m} V^{N}{}_{n} V^{P}{}_{p} \brrho \gamma^{mnp} \rho + i f_{MNP} \bar{V}^{M}{}_{\bar{m}} V^{N}{}_{n} V^{P}{}_{p} \brpsi^{\bar{m}}\gamma^{np} \rho 
\\&
+i \frac{1}{3} f_{MNP} V^{M}{}_{m} V^{N}{}_{n} V^{P}{}_{p} \brpsi^{\bar{n}}\gamma^{mnp} \psi_{\brn} + 4i f_{MNP}V^{M}{}_{m} \bar{V}^{N}{}_{\bar{n}} \bar{V}^{P}{}_{\bar{p}} \brpsi^{\bar{n}}\gamma^{m} \psi^{\bar{p}}
 \Big]\,.
\end{array}
\end{equation}
The half-maximal supersymmetric gauged DFT (\ref{SGDFT}) is invariant under the following SUSY transformations up to leading order in fermions,
\be\ba{l}
\delta d = -i\half \brvare \rho\,,
\\
\delta V_{Mm} = -i \brV_{M}{}^{\brq} \brvare \gamma_{m} \psi_{\brq}\,,
\\
\delta \brV_{M\brm} = i \brV_{M}{}^{q} \brvare \gamma_{q} \psi_{\brm}\,,
\\
\delta \rho = - \gamma^{m} \cD_{m} \varepsilon\,,
\\
\delta \psi_{\brm} = \cD_{\brm} \varepsilon\,,
\ea\ee
where SUSY parameter $\varepsilon$ is a  $\Spin(1,9)$ spinor with positive chirality,
\be
\gamma^{(11)} \varepsilon = \varepsilon\,.
\ee
The supersymmetry variation of the gauged DFT action (\ref{SGDFT}) up to leading order in fermions is given by
\be\ba{ll}
\delta \cL_{\rm SGDFT} = e^{-2d} \Big[& - 4 \delta d P^{MP} P^{NQ} S_{MNPQ} + 4 \delta P^{MP} P^{NQ} S_{MNPQ}  
\\
&+ 8i \brrho \left(  \gamma^{m} \cD_{m} \delta \rho + \cD_{\brm} \delta\psi^{\brm} \right)- 8 i \brpsi^{\brm} \left(\gamma^{m} \cD_{m}\delta\psi_{\brm} + \cD_{\brm} \delta \rho\right)
\Big] .
\ea\ee
We can then check the supersymmetry invariance of the action (\ref{SGDFT}) by using the following identities, 
\be\ba{rl}
\gamma^{m}\gamma^{n} \cD_{m} \cD_{n} \varepsilon+ \cD^{\brm} \cD_{\brm}\varepsilon =& - \tfrac{1}{4}\varepsilon P^{MN} P^{PQ} S_{MQNP}\,,
\\
\gamma^{n} \comm{\cD_{\brm}}{\cD_{n}} \varepsilon =& \brV^{M}_{\brm} V^{N}{}_{n} P^{PQ} S_{MPNQ} \gamma^{n}\varepsilon\,.
\ea\ee

What is extraordinary is how simple the action (\ref{SGDFT}) is in terms of these doubled gauged geometric quantities.

%%%%%%%%%%%%%%%%
\section{Generalized Scherk-Schwarz reduced DFT as a gauged DFT} 

In this section we show how using this formalism, the gauged double field theory can be obtained from the generalised Scherk-Schwarz reduction from the higher dimensional ungauged double field theory. Let hatted indices $\hM, \hN, \hP, \cdots$ represent $\ODD$ vector indices in parent ungauged DFT  and $M,N,P,\cdots$ represent $\ODD$ vector indices in gauged double field theory. We divide the $2D$ dimensional doubled spacetime coordinates  $\hat{\mathbb{X}}^{\hM}$ into $2d$-dimensional non-compact space coordinatees $\mathbb{X}^{M}$ and $2n$-dimensional compact space coordinates $\mathbb{Y}^{I}$. If we introduce a twist matrix $U_{\hat{M}}{}^{M} (\mathbb{Y})$, the Scherk-Schwarz reduction is realised as
\be\ba{rl}
\hV_{\hM}(\mathbb{X}, \mathbb{Y}) =& U_{\hM}{}^{M}(\mathbb{Y}) V_{M}(\mathbb{X})\,,
\\
\hd(\mathbb{X}, \mathbb{Y}) =& d(\mathbb{X}) + \lambda(\mathbb{Y})  
\ea\label{ssred}\ee
where $\hV_{\hM}$ is an $\ODD$ vector that depends on the noncompact directions only and $e^{-2\hd}$ is a tensor density. 

Once we have this, the generalised Scherk-Schwarz reduction of the parent DFT connection (\ref{oldconn}) is realised from substitution of Scherk-Schwarz ansatz (\ref{ssred}) into the definition of parent DFT connection (\ref{oldconn}),
\be
\hat{\Gamma}_{\hP\hM\hN} (\hP, \hbP, \hd )  = 
 U_{\hP}{}^{P} U_{\hM}{}^{M} U_{\hN}{}^{N} \hat{\Gamma}_{PMN}(P,\brP,d) \,, \label{SSconan}
\ee
where
\be
\ba{ll}
\hat{\Gamma}_{PMN} &= \Gammao_{PMN}  
 + (P_{[M}{}^{Q} P_{N]}{}^{R} + \brP_{[M}{}^{Q} \brP_{N]}{}^{R} ) f_{PQR}  - (U^{-1})_{M}{}^{\hQ} \partial_{P} U_{\hQ N} 
\\ 
 & ~~~
 - 2 (\cP+\bar{\cP})_{PMN}{}^{QRS} (U^{-1})_{S}{}^{\hT}\partial_{R} U_{\hT Q}  
- \tfrac{2}{D-1} (\brP_{P[M} \brP_{N]}{}^{Q} + P_{P[M} P_{N]}{}^{Q}) f_{Q} \,.
\ea \label{redconn}\ee
As before, $\Gamma^{0}{}_{PMN}$ is the connection for ungauged DFT. Here $f_{MNP}$ and $f_{M}$ are defined by
\be\ba{rl}
f_{MNP} =& 3\eta_{Q[M} (U^{-1})_{N}{}^{\hN} (U^{-1})_{P]}{}^{\hM} \partial_{\hN}U_{\hM}{}^{Q}\,,
\\ f_{M} =& \partial_{\hM} (U^{-1})_{M}{}^{\hM} -2  (U^{-1})_{M}{}^{\hM} \partial_{\hM}\lambda\,.
\ea\ee
The reader may ask whether (\ref{SSconan}) is covariant. As a connection it is of course not. We just follow the generalised Scherk-Schwarz ansatz and from this connection the guage connections will also emerge.   The $f_{MNP}$ can be identified with the structure constant in twisted generalised Lie derivative (\ref{symm}) as shown in \cite{Grana:2012rr}.  Also, for consistency, we need to set $f_{A}=0$ just as in \cite{Grana:2012rr}.

It is important to compare the reduced connection with the gauged DFT connection in (\ref{conn}).   If we calculate the difference, then we have
\be\ba{ll}
\left(\hat{\Gamma}- \Gamma\right)_{MNP} = - (U^{-1})_{N}{}^{\hQ} \partial_{M} U_{\hQ P} 
 + \left(\cP+\bar{\cP}\right)_{MNP}{}^{QRS} \left( \tfrac{2}{3} f_{QRS} - 2 (U^{-1})_{S}{}^{\hT}\partial_{R} U_{\hT Q} \right) \,,
\ea\ee
Note, $U_{\hQ P}$ does not obey strong section condition!
On the right hand side, the last term is removed after contraction with a projection operator and so does not contribute to the fully covariant quantities. The first term, $- (U^{-1})_{N}{}^{\hQ} \partial_{M} U_{\hQ P} $ however does contribute. This is the difference between reduced parent DFT connection $\hat{\Gamma}_{PMN}$ in (\ref{redconn}) and gauged DFT connection $\Gamma_{PMN}$ in (\ref{conn}) and it shows the origin in terms of the reduced connection of the additional term in the action that appeared in  \cite{Grana:2012rr}.

As discussed before, the gauged DFT action should be independent of $\mathbb{Y}^{I}$ or $U_{\hM}{}^{M}(\mathbb{Y})$ so that,
\be
\cS_{\rm GDFT} = \int \rd^{2d}\mathbb{X} \,\cL_{\rm GDFT} \left[P, \brP, d, f\right]\,.
\ee
 However, if we carry out a generalised Scherk-Schwarz reduction on the parent DFT action,  the reduced action is written in terms of  $\hat{\Gamma}_{PMN}$ having explicit twist matrix dependence, $U_{\hM}{}^{M}(\mathbb{Y})$, 
\be
\cS_{\rm red} = \int \rd^{2n} \mathbb{Y} \int \rd^{2d} \mathbb{X} \, \cL_{\rm red} \left[\hat{\Gamma}(P, \brP, d, f, U)\right]\,.
\label{redaction}\ee 
To get a $U_{\hM}{}^{M}(\mathbb{Y})$ independent action, an additional term should be added which compensates the $U_{\hM}{}^{M}(\mathbb{Y})$ dependence of the reduced action (\ref{redaction}).  The additional term is given by
\be\ba{ll}
\cL_{\rm GDFT} - \cL_{\rm red} &= e^{-2d} \left( P^{MN} P^{PQ} U_{\hM M} U_{\hN N} \partial_{R}(U^{-1})_{P}{}^{\hM}  \partial^{R}(U^{-1})_{Q}{}^{\hN}\right) \,,
\\
&= e^{-2d} \left( \half \partial_{M} (U^{-1})^{N \hQ}\partial^{M} U_{\hQ}{}^{P} \cH_{NP} \right) \,.
\ea\ee
This term exactly reproduces the additional term in (2.4) of  \cite{Grana:2012rr}, 
\be
\half \partial_{\hM} \varepsilon^{a}{}_{\hP} \partial^{\hM} \varepsilon^{b}{}_{\hQ} S_{ab} \eta^{\hP\hQ}= \half \partial_{M} (U^{-1})^{N \hQ}\partial^{M} U_{\hQ}{}^{P} \cH_{NP} \,.
\ee

It is worthwhile to compare how supersymmetry works. The difference of spin-connections is given by
\be\ba{ll}
\left(\hat{\Phi} - \Phi\right)_{Mmn} = 
  V^{N}{}_{m} V^{P}{}_{n}  \cP_{MNP}{}^{QRS} \left( \tfrac{2}{3} f_{QRS} -2 (U^{-1})_{S}{}^{\hT}\partial_{R} U_{\hT Q} \right) \,,
 \\
 \left(\hat{\brPhi} - \brPhi\right)_{M\brm\brn} = 
 \brV^{N}{}_{\brm} \brV^{P}{}_{\brn} \bar{\cP}_{MNP}{}^{QRS} \left( \tfrac{2}{3} f_{QRS} - 2 (U^{-1})_{S}{}^{\hT} \partial_{R} U_{\hT Q}  \right) \,.
\ea\ee
When we then insert these into the supersymmetry transformations, then these differences vanish for the semi-covariant derivatives. This shows that Killing spinor equations for gauged DFT and generalised Scherk-Schwarz reduced DFT are exactly same. This is remarkable that the supersymmetry remains identical between the gauged and Scherk-Schwarz reduced cases even though there are differences in the connections that require correction terms in the action.

\section{Killing Spinor Equations}

In the context of ordinary supergravity, the Killing spinor equations have proven very useful in finding solutions to the equations of motion that preserve some fraction of supersymmetry. Essentially they reduce the supergravity equations to be first order in derivatives and often when combined with a suitable ansatz for the metric and fields will lead to linear equations. Importantly, the Scherk-Schwarz factorisation ansatz for double field theory allows dependence on both the usual coordinates and their duals simultaneously. Finding actual solutions that obey this ansatz is much more difficult than finding solutions that obey the strong constraint. Obviously from the connection to gauged field supergravity there is a clear interpretation of some of these solutions. However, one might wish to just try and seek solutions in double field theory with a Scherk-Schwarz ansatz and interpret this as a double field theory geometry. This then involves solving the double field theory equations of motion. Obviously this is a hard problem since the double field theory equations of motion are as hard to solve as Einstein's equations.

The natural thing is to then use the Killing spinor equations with a Scherk-Schwarz factorisation ansatz (reduction). One can then seek solutions in the full double field theory preserving different fractions of supersymmetry but that obey the Scherk-Schwarz ansatz. In what follows we write down the Killing spinor equations with the Scherk-Schwarz anstaz.

The Killing spinor equations for the higher dimensional ungauged DFT were given already in \cite{Jeon:2011sq, Jeon:2012hp}
\be\ba{ll}
\delta \hat{\rho}(\mathbb{X}, \mathbb{Y}) = - \gamma^{\hm}  \hat{\cD}_{\hm} \hat{\varepsilon}(\mathbb{X}, \mathbb{Y}) = 0\,,
\\ 
 \delta \hat{\psi}_{\hbm}(\mathbb{X}, \mathbb{Y}) = \hat{\cD}_{\hbm} \hat{\varepsilon}(\mathbb{X}, \mathbb{Y}) = 0\,.
\ea\ee
Since the spinors are all $\ODD$ scalars, the generalised Scherk-Schwarz ansatz for the spinors is trivial,
\be
\hat\rho(\mathbb{X}, \mathbb{Y}) = \rho(\mathbb{X})\,,  ~~~~~~~~~~ \hat{\psi}_{\hat{\brm}}(\mathbb{X}, \mathbb{Y}) = \psi_{\hat{\brm}}(\mathbb{X})\,, ~~~~~~~~~~ \hat\varepsilon(\mathbb{X}, \mathbb{Y}) = \varepsilon(\mathbb{X})\,.
\ee
Althought the spinors reduce trivially the spin conncetions do not and one must take care to reduce them. We list the reduction of the relevent spin connections that are needed for the Killing spinor equations in the appendix. Inserting these generalised Scherk-Schwarz reduced spin connections into to the Killing spinor equations, produces:
\be\ba{ll}
-\gamma^{\hm} \hat{\cD}_{\hm}\hat{\varepsilon} =& -  \gamma^{m} \cD_{m} \varepsilon + \tfrac{3}{4}V^{M}{}_{m} V^{N}{}_{n} V^{P}{}_{p} \left(\cA_{[M} \partial_{N} \cA_{P]} -\tfrac{1}{3} f^{ABC} \cA_{MA} \cA_{NB} \cA_{PC}\right)\gamma^{np}\varepsilon 
\\&
+ \tfrac{1}{4} (\cF_{MN})_{A} V^{M}{}_{m}V^{N}{}_{n} V^{A}{}_{a} \gamma^{mna}\varepsilon - \tfrac{1}{4}V^{M}{}_{m} V^{A}{}_{a} D_{M}V_{Ab} \gamma^{mab}\varepsilon 
\\&
- \tfrac{1}{4}V^{A}{}_{a} V^{B}{}_{b} V^{C}{}_{c} f_{ABC} \gamma^{abc}\varepsilon = 0\,,
\ea\ee
for dilatino and
\be\ba{ll}
\hat{\cD}_{\brm} \hat\varepsilon =& \cD_{\brm} \varepsilon - \tfrac{3}{4}\brV^{M}{}_{\brm} V^{N}{}_{n} V^{P}{}_{p} \left(\cA_{[M} \partial_{N} \cA_{P]} -\frac{1}{3} f^{ABC} \cA_{MA} \cA_{NB} \cA_{PC}\right)\gamma^{np}\varepsilon 
\\&
- \half \brV^{M}{}_{\brm} V^{N}{}_{n} V^{A}{}_{a} (\cF_{MN})_{A} \gamma^{na} \varepsilon + \frac{1}{4} \brV^{M}{}_{\brm} V^{A}{}_{a} D_{M} V_{A b} \gamma^{ab}\varepsilon = 0\,,
\\
\hat{\cD}_{\bra} \hat\varepsilon = &- \frac{1}{4} \brV^{A}{}_{\bra} V^{M}{}_{m} V^{N}{}_{n} (\cF_{MN})_{A} \gamma^{mn} \varepsilon - \half \brV^{A}{}_{\bra} V^{M}{}_{m} V^{B}{}_{b} D_{M}P_{AB} \gamma^{mb} \varepsilon
\\&
+\tfrac{1}{4} \brV^{A}{}_{\bra} V^{B}{}_{b} V^{C}{}_{c} f_{ABC} \gamma^{bc}\varepsilon = 0\,
\ea\ee
for the gravitino. 

A great deal of insight has been achieved through the analysis of the usual Killing spinor equations. It would be very interesting to investigate these equations that come from double field theory in detail. Hopefully one could obtain results along the lines as of the G-structure geometric spinor approach \cite{Gauntlett:2003fk}.
We leave this for future work. 

\bigskip
\bigskip

\noindent{{\textbf{Acknowledgements.}}   We would like to thank Jeong-Hyuck Park and Malcolm Perry for helpful discussions and comments. K. L. is  supported by the National Research Foundation of Korea(NRF) and  the Ministry of Education, Science and Technology with the Grant No. 2005-0049409 (CQUeST) and  No. 2012R1A6A3A03040350.  DSB is partially supported by STFC consolidated grant ST/J000469/1. DSB is also grateful for DAMTP in Cambridge for continuous hospitality.

\appendix
%%%%%%%%%%%%%%%%
\section{Decomposition of compact and non-compact directions}
In this section we describe the Scherk-Schwarz reduction in detail. This is all covered in \cite{Aldazabal:2011nj} but we have repeated it here to make conventions clear and provide a quick reference.

\subsection{Reduction conventions}
Now we consider explicit breaking of $\ODD$ symmetry into $\Onn$ subgroup.
We decompose $D$-dimensional total spacetime into $n$-dimensional compact and $d$-dimensional non-compact direction. All hatted indices represent quantities defined on total spacetime. 
\begin{enumerate}
\item Total spacetime :
\begin{itemize}
\item $\hM,\hN, \cdots$ : $\ODD$ vector indices,
\item $\hmu,\hnu, \cdots$ : $D$-dimensional vector indices,
\item $\hm, \hn, \cdots$ : Local $\SpinD$ vector indices,
\item $\hbm,\hbn, \cdots$ : Local $\oSpinD$ vector indices.
\end{itemize}
\item Non-compact direction :
\begin{itemize}
\item $M,N,\cdots $ : $\Odd$ vector indices,
\item $\mu,\nu, \cdots$ : $d$-dimensional vector indices,
\item $m, n, \cdots$ : Local $\Spind$ vector indices,
\item $\brm,\brn, \cdots$ : Local $\oSpind$ vector indices.
\end{itemize}
\item Compact direction :
\begin{itemize}
\item $I, J,\cdots $ : $\Onn$ vector indices,
\item $A, B,\cdots $ : Gauge indices,
\item $\alpha,\beta, \cdots$ : $n$-dimensional vector indices,
\item $a, b, \cdots$ : Local $\Spinn$ vector indices,
\item $\bra,\brb, \cdots$ : Local $\oSpinn$ vector indices.
\end{itemize}
\end{enumerate}
Therefore doubled spacetime coordinates $\hat{\mathbb{X}}^{\hM} = \{\hx_{\hmu}, \hx^{\hnu}\}$ are decomposed into 
\be
\hat{\mathbb{X}}^{\hM} = \{\mathbb{X}^{M}, \mathbb{Y}^{I}\}\,,
\ee
where $\mathbb{X}^{M}= \{x_{\mu}, x^{\nu}\}$ is non-compact direction doubled coordinate and $\mathbb{Y}^{I} = \{ y_{\alpha}, y^{\beta} \}$ is compact direction doubled coordinate.
%%%%%%%%%%
\subsection{Reduction of ordinary supergravity}
The ordinary Scherk-Schwarz reduction of two copies of the $D$-dimensional viellein is given by
\be
\he_{\hmu}{}^{\hm} =\bpm e_{\mu}{}^{m} & A_{\mu}{}^{\alpha} \Phi_{\alpha}{}^{a}
\\0 & \Phi_{\alpha}{}^{a} \epm\,, ~~~~~~~~~~~ (\he^{-1})_{\hm}{}^{\hmu} = \bpm (e^{-1})_{m}{}^{\mu} & - (e^{-1})_{m}{}^{\mu} A_{\mu}{}^{\alpha} \\ 0 & (\Phi^{-1})_{a}{}^{\alpha} \epm\,,
\label{ssansatzV}
\ee
and
\be
\hbe_{\hmu}{}^{\hbm} =\bpm \bre_{\mu}{}^{\brm} & A_{\mu}{}^{\alpha} \brPhi_{\alpha}{}^{\bra}
\\0 & \brPhi_{\alpha}{}^{\bra} \epm\,, ~~~~~~~~~~~ (\hbe^{-1})_{\hbm}{}^{\hmu} = \bpm (\bre^{-1})_{\brm}{}^{\mu} & -(\bre^{-1})_{\brm}{}^{\mu}  A_{\mu}{}^{\alpha} \\ 0 & (\brPhi^{-1})_{\bra}{}^{\alpha} \epm\,.
\label{ssansatzbrV}
\ee
where 
\be
\ba{l}
e_{\mu}{}^{m} e_{\nu m} = - \bre_{\mu}{}^{\brm} \bre_{\nu \brm} = g_{\mu\nu}\,,
\\
\Phi_{\alpha}{}^{a} \Phi_{\beta a} = - \brPhi_{\alpha}{}^{\bra} \brPhi_{\beta \bra} = g_{\alpha \beta}
\ea
\ee
The reduction of Kalb-Ramond field is then:
\be
\hB_{\hmu\hnu} = \bpm \hB_{\mu \nu} & \hB_{\mu \beta} \\ \hB_{\alpha \nu} & \hB_{\alpha \beta} \epm := \bpm B_{\mu \nu} + \half (A_{\mu}{}^{\alpha} B_{\alpha \nu} - A_{\nu}{}^{\alpha} B_{\alpha \mu}) + A_{\mu}{}^{\alpha} A_{\nu}{}^{\beta} B_{\alpha \beta} &  B_{\mu \beta} + A_{\mu}{}^{\alpha} B_{\alpha\beta} \\ B_{\alpha \nu} + B_{\alpha \beta} A_{\nu}{}^{\beta} & B_{\alpha \beta} \epm\,.
\ee
%%%%%
\subsection{Scherk-Schwarz reduction of Double Field Theory}
The double vielbein of the total space is parametrized by
\be
\hV_{\hM}{}^{\hm} = \frac{1}{\sqrt{2}}
\begin{pmatrix}(\he^{-1})^{\hm\hmu}\\ (\hB+\he)_{\hmu}{}^{\hm}
\end{pmatrix}\,,~~~~~~~~~~ \hbV_{M}{}^{\hbm} = \frac{1}{\sqrt{2}} \bpm (\bre^{-1})^{\hbm \hmu}\\ (\hbB+\bre)_{\mu}{}^{\hbm}\epm 
\ee
where $\hB_{\mu}{}^{\hm} = \hB_{\hmu\hnu} (\he^{-1})^{\hm\hnu}$ and $\hbB_{\hmu}{}^{\hbm} = \hB_{\hmu \hnu} (\hbe^{-1})^{\hbm \hnu}$.
\\~\\
Let us now consider the Scherk-Schwarz reduction ansatz for the double vielbein, $\hV_{\hM}{}^{\hm}$ and $\hbV_{\hM}{}^{\hbm}$
\be
\hV_{\hM}{}^{\hm}(\mathbb{X}, \mathbb{Y}) \longrightarrow \bpm \hV_{M}{}^{m}(\mathbb{X}) & \hV_{M}{}^{a}(\mathbb{X}) \\ U_{I}{}^{A}(\mathbb{Y}) \hV_{A}{}^{m}(\mathbb{X}) & U_{I}{}^{A}(\mathbb{Y}) \hV_{A}{}^{a}(\mathbb{X}) \epm\,,
\label{redV}
\ee
and 
\be
\hbV_{\hM}{}^{\hbm}(\mathbb{X}, \mathbb{Y}) \longrightarrow \bpm \hbV_{M}{}^{\brm}(\mathbb{X}) & \hbV_{M}{}^{\bra}(\mathbb{X}) \\ U_{I}{}^{A}(\mathbb{Y}) \hbV_{A}{}^{\brm}(\mathbb{X}) & U_{I}{}^{A}(\mathbb{Y}) \hbV_{A}{}^{\bra}(\mathbb{X}) \epm\,,
\label{redbrV}
\ee
where $U_{I}{}^{A}(\mathbb{Y})$ is a generalised twist matrix. Finally, the Scherk-Schwarz ansatz of dialton is given by 
\be
\hd(\mathbb{X}, \mathbb{Y}) = d(\mathbb{X}) + \lambda(\mathbb{Y})\,.
\ee
Importantly, when using the the double vielbein, it is not possible to choose a upper triangular form as (\ref{ssansatzV}) and (\ref{ssansatzbrV}), since the local Lorentz group is not sufficient. For example, unbared double vielbein $\hV_{\hM}{}^{\hm}$ has only $\SpinD$ local Lorentz symmetry instead of $\Spin(D,D)$.

Each component of reduced double vielbeins in eq. (\ref{redV}) and (\ref{redbrV}) can be written as follows:
\be
\ba{ll}
\hV_{M}{}^{m} = V_{M}{}^{m} -\half \cA_{M}{}^{A} \cA^{N}{}_{A} V_{N}{}^{m}\,, ~~~~~~&~~~~~~ \hV_{M}{}^{a} = -\cA_{M}{}^{A} V_{A}{}^{a}\,,
\\ 
\hV_{A}{}^{m} = \cA^{M}{}_{A} \hV_{M}{}^{m} =\cA^{M}{}_{A}V_{M}{}^{m} \,,~~~~~~&~~~~~~ \hV_{A}{}^{a} = V_{A}{}^{a}\,,
\ea
\label{ssredV}
\ee
and 
\be
\ba{ll}
\hbV_{M}{}^{\brm} = \brV_{M}{}^{\brm} -\half \cA_{M}{}^{A} \cA^{N}{}_{A} \brV_{N}{}^{\brm}\,, ~~~~~~&~~~~~~ \hbV_{M}{}^{\bra} = -\cA_{M}{}^{A} \brV_{A}{}^{\bra}\,,
\\
\hbV_{A}{}^{\brm} = \cA^{M}{}_{A} \hbV_{M}{}^{\brm} =\cA^{M}{}_{A} \brV_{M}{}^{\brm} \,,~~~~~~&~~~~~~ \hbV_{A}{}^{\bra} = \brV_{A}{}^{\bra}\,,
\ea
\label{ssredbrV}
\ee
where $\cA_{M}{}^{A}$ is an unified gauge field,
\be
\cA_{\mu}{}^{A} = \bpm -A_{\mu}{}^{\alpha} \\ -\hB_{\alpha \beta} A_{\mu}{}^{\beta} + \hB_{\alpha \mu} \epm := \bpm -A_{\mu}{}^{\alpha} \\ B_{\alpha \mu} \epm \,,~~~~~ \cA^{\mu A} = 0\,.
\ee
Here, $V_{M}{}^{m}$, $\brV_{M}{}^{\brm}$ are $d$-dimensional double vielbeins parametrized as
\be
\ba{ll}
V_{M}{}^{m} = \frac{1}{\sqrt{2}}
\bpm(e^{-1})^{m \mu}\\ (B+e)_{\mu}{}^{m} \epm\,,~~~~ &~~~\brV_{M}{}^{\brm} = \frac{1}{\sqrt{2}}
\bpm(\bre^{-1})^{\brm \mu}\\ (\brB+\bre)_{\mu}{}^{\brm} \epm\,,
\ea
\ee
and $V_{A}{}^{a}$, $\brV_{A}{}^{\bra}$ are $n$-dimensional double vielbeins parametrized as
\be
\ba{ll}
V_{A}{}^{a} = \frac{1}{\sqrt{2}} \bpm(e^{-1})^{a \alpha}\\ (B+e)_{\alpha}{}^{a} \epm\,,
~~~~ &~~~
\brV_{A}{}^{\bra} = \frac{1}{\sqrt{2}} \bpm(\bre^{-1})^{\bra \alpha}\\ (\brB+\bre)_{\alpha}{}^{\bra} \epm\,.
\ea
\ee
One can find the transformation laws for various fields by substituting the reduced vielbeins defined in eq. (\ref{ssredV}) and (\ref{ssredbrV}) into double-gauge transformation(or generalised Lie derivative),
\be
\hcL_{\hX} \hV_{\hM}{}^{\hm} = \hX^{\hN} \partial_{\hN} \hV_{\hM}{}^{\hm} + (\hat{\partial}_{\hM} \hX^{\hN} - \hat{\partial}^{\hN} \hX_{\hM}) \hV_{\hN}{}^{\hm}\,.
\ee
To examine the symmetry transformations of the reduced fields, we should decompose the double gauge parameter $\hX^{\hM}$ as before, 
\be
\hX^{\hM}(\mathbb{X}, \mathbb{Y}) = \bpm X^{M}(\mathbb{X})\\ (U^{-1})_{A}{}^{I}(\mathbb{Y}) Y^{A}(\mathbb{X})\epm\,.
\ee
We then interpret $X^{M}$ as a generalised Lie derivative parameter and $Y^{A}$ as a gauge symmetry parameter.\\
The symmetry transformation of each of the reduced fields is given by
\be
\ba{lll}
&\delta V_{M m} = \hcL_{X} V_{M m} - \cA_{[M}{}^{A} \partial_{N]} Y_{A} V^{N m} \,, ~~~&~~~ \delta V_{A a} = \hcL_{X} V_{A a} - f_{AB}{}^{C} Y^{B} V_{Ca}\,,
\\
&\delta \brV_{M \brm} = \hcL_{X} \brV_{M \brm} - \cA_{[M}{}^{A} \partial_{N]} Y_{A} \brV^{N \brm}\,, ~~~&~~~ \delta \brV_{A \bra} = \hcL_{X} \brV_{A \bra} - f_{AB}{}^{C} Y^{B} \brV_{C\bra}\,,
\\
&\delta e^{-2 d} = \hcL_{X} e^{-2d} + f_{A} Y^{A} e^{-2d}\,, ~~~&~~~~ 
\\
&\delta \cA_{M}{}^{A} = \hcL_{X} \cA_{M}{}^{A} - \partial_{M}Y^{A} + f^{A}{}_{BC}\cA_{M}{}^{B} Y^{C} \,,&
\ea
\ee
where
\be
f_{ABC} = 3\eta_{D[A} (U^{-1})_{C}{}^{I} \partial_{B]}U_{I}{}^{D}\,,~~~~~~~~~ f_{A} = U_{I}{}^{B} \partial_{B} (U^{-1})_{A}{}^{I} -2 \partial_{A}\lambda\,.
\ee
Since projection operators can be written in terms of double vielbein, 
\be
\hV_{\hM}{}^{\hm} \hV_{\hN \hm} = \hP_{\hM \hN}\,, ~~~~~~~~~~~~ \hbV_{\hM}{}^{\hbm} \hbV_{\hN \hbm} = \hbP_{\hM \hN}\,,
\ee
the Scherk-Schwarz reduction of the projection operators may be easily obtained using the reduction of the double vielbeins from (\ref{ssredV}) and (\ref{ssredbrV}), which yields
\be
\hP_{\hM \hN}(\mathbb{X},\mathbb{Y}) = \bpm \hP_{MN}(\mathbb{X})& U_{J}{}^{B}(\mathbb{Y}) \hP_{MB}(\mathbb{X})\\ U_{I}{}^{A}(\mathbb{Y}) \hP_{A N}(\mathbb{X}) & U_{I}{}^{A}(\mathbb{Y}) U_{J}{}^{B}(\mathbb{Y}) \hP_{A B}(\mathbb{X}) \epm\,,
\ee
and 
\be
\hbP_{\hM \hN}(\mathbb{X},\mathbb{Y}) = \bpm \hbP_{MN}(\mathbb{X})& U_{J}{}^{B}(\mathbb{Y}) \hbP_{MB}(\mathbb{X})\\ U_{I}{}^{A}(\mathbb{Y}) \hbP_{A N}(\mathbb{X}) & U_{I}{}^{A}(\mathbb{Y}) U_{J}{}^{B}(\mathbb{Y}) \hbP_{A B}(\mathbb{X}) \epm\,.
\ee
Each component of $\hP_{\hM\hN}$ is 
\be
\ba{ll}
\hP_{MN}=P_{MN} - \cA_{(M}{}^{A} \cA^{P}{}_{|A} P_{P|N)} +\frac{1}{4} \cA_{M}{}^{A}\cA^{P}{}_{A} \cA_{N}{}^{B} \cA^{Q}{}_{B} P_{PQ} + \cA_{M}{}^{A} \cA_{N}{}^{B}  P_{A B}\,,
\\
\hP_{MA} = \hP_{A M} = (\cJ_{M P} - \half \cA_{M}{}^{B} \cA_{P B}) P^{P N} \cA_{N A} - \cA_{M}{}^{B} P_{A B}\,,
\\
\hP_{A B} = P_{A B} + \cA^{M}{}_{A} \cA^{N}{}_{B} P_{MN} \,,
\ea
\ee
where 
\be
P_{MN} = V_{M}{}^{m} V_{N m}\,, ~~~~~~~~~~~ P_{AB} = V_{A}{}^{a} V_{B a}\,.
\ee
Similarly, each component of $\hbP_{\hM\hN}$ is given by
\be
\ba{ll}
\hbP_{MN} = \brP_{MN} - \cA_{(M}{}^{A} \cA^{P}{}_{|A} \brP_{P|N)} + \frac{1}{4} \cA_{M}{}^{A}\cA^{P}{}_{A} \cA_{N}{}^{B} \cA^{Q}{}_{B} \brP_{PQ} + \cA_{M}{}^{A} \cA_{N}{}^{B}  \brP_{A B}\,,
\\
\hbP_{MA} = \hbP_{A M} = (\cJ_{M P} -\half \cA_{M}{}^{B} \cA_{P B}) \brP^{P N} \cA_{N A} - \cA_{M}{}^{B} \brP_{A B}\,,
\\
\hbP_{A B} = \brP_{A B} + \cA^{M}{}_{A} \cA^{N}{}_{B} \brP_{MN}\,,
\ea
\ee
where 
\be
\brP_{MN} = \brV_{M}{}^{\brm} \brV_{N \brm}\,, ~~~~~~~~~~~ \brP_{AB} = \brV_{A}{}^{\bra} \brV_{B \bra}\,.
\ee

If we apply these reduction conventions, the bosonic part of the half-maximal supersymmetric gauged double field theory action (\ref{SGDFT}) is reduced to
\be
\ba{ll}
\cL_{\rm SGDFT} =  e^{-2d} &\left( 2P^{MP}P^{NQ} \So_{MNPQ} + \half \cH^{MN} \cH^{PQ} \omega_{MPR}\partial_{N}\cH_{QP} -\frac{1}{12} \cH^{MN} \cH^{PQ} \cH_{RS} \omega_{MPR} \omega_{NQS}\right.
\\&\left. 
 + \frac{1}{8} \cH^{MN} D_{M}\cH_{AB} D_{N}\cH^{AB} -\frac{1}{4} \cH^{MN} \cH^{PQ} \cH_{AB} (F_{MP})^{A} (F_{NQ})^{B} \right.
\\ &\left. 
-\tfrac{1}{12} \cH^{MN} \cH^{PQ} \cH^{RS} f_{MPR} f_{NQS} + \tfrac{1}{4} \cH^{PQ} f_{MNP} f^{MN}{}_{Q} + \tfrac{1}{6} f_{MNP} f^{MNP} \right) \,,
\ea\label{redGDFT}
\ee
where $\omega_{MNP}$ is Chern-Simon 3-form,
\be
\omega_{MNP} := 3 \cA_{[M}{}^{A} \partial_{N} \cA_{P]A} - \cA_{M}{}^{A} \cA_{N}{}^{B} \cA_{P}{}^{C} f_{A B C} \,,
\ee
and covariant derivative, $D_{M}$, for the gauge transformation generated by $Y^{A}$ and field strength, $\cF_{M N}$, for the unified gauge field, $\cA_{M}{}^{A}$ , are defined as 
\be
\ba{l}
D_{M} V_{A} := \partial_{M} V_{A} - f_{A B C} \cA_{M}{}^{B} V^{C}\,,
\\
(\cF_{MN})^{A} := \partial_{M}\cA_{N}{}^{A} - \partial_{N}\cA_{M}{}^{A} - f^{A}{}_{BC} \cA_{M}{}^{B} \cA_{N}{}^{C}\,.
\ea
\ee
One has thus shown that the reduced action (\ref{redGDFT}) is exactly same as the half-maximal gauged supergravity action in \cite{Aldazabal:2011nj}.

\section{Reduction of spin-connections \label{appendix}}
Though the spin connections, $\hPhi_{\hM \hm \hn}$ and $\hbPhi_{\hM \hbm\hbn}$, are not covariant under the double-gauge transformation, the following terms are covariant, 
\be
\ba{l}
\hPhi_{[\hm\hn\hp]}\,, ~~~~~~~~ \hPhi^{\hm}{}_{\hm\hp}\,, ~~~~~~~~ \hPhi_{\hbp\hm\hn} \,,\\
\hbPhi_{[\hbm\hbn\hbp]}\,, ~~~~~~~~ \hbPhi^{\hbm}{}_{\hbm\hbp}\,, ~~~~~~~~ \hbPhi_{\hp\hbm\hbn} \,,
\ea
\ee
where $\hPhi_{\hm \hn \hp} := \hV^{\hM}{}_{\hm} \hPhi_{\hM \hn \hp}$ and  $\hPhi_{\hbm \hn \hp} := \hbV^{\hM}{}_{\hbm} \hPhi_{\hM \hn \hp}$, etc. 

Reduction of the covariant combinations of $\hPhi_{\hM \hm \hn}$ yields:
\be
\ba{cl}
\hPhi_{[mnp]} &= \Phi_{[mnp]} -3  V^{M}{}_{m} V^{N}{}_{n} V^{P}{}_{p} \left(\cA_{[P}{}^{A} \partial_{M} \cA_{N] A} -\frac{1}{3} f^{ABC}\cA_{PA} \cA_{MB} \cA_{NC}\right)\,,
\\
\hPhi_{[mna]} &= - V^{M}{}_{m} V^{N}{}_{n} V^{A}{}_{a} (\cF_{MN})_{A}\,,
\\
\hPhi_{[mab]} &= V^{M}{}_{m} V^{A}{}_{a} D_{M}V_{A b}\,,
\\
\hPhi_{{[abc]}} &= V^{A}{}_{a}V^{B}{}_{b}V^{C}{}_{c} f_{ABC}\,,
\\
\hPhi^{\hn}{}_{\hn m} &= \Phi^{n}{}_{n m} + V^{M}{}_{m} \cA_{M}{}^{A} f_{A} \,,
\\
\hPhi^{\hn}{}_{\hn a} &= V^{A}{}_{a} f_{A}\,,
\\
\hPhi_{\brp mn} &= \Phi_{\brp m n} - 3 \brV^{P}{}_{\brp} V^{M}{}_{m} V^{N}{}_{n} \left(\cA_{[P}{}^{A} \partial_{M} \cA_{N] A} -\frac{1}{3} f^{ABC}\cA_{PA} \cA_{MB} \cA_{NC}\right)\,,
\\
\hPhi_{\bra m n} &= - \brV^{A}{}_{\bra} V^{M}{}_{m} V^{N}{}_{n} (\cF_{MN})_{A}\,,
\\
\hPhi_{\brp m a} &= - \brV^{M}{}_{\brp} V^{N}{}_{m} V^{A}{}_{a} (\cF_{MN})_{A}\,, 
\\
\hPhi_{\brp a b} &= \brV^{M}{}_{\brp} V^{A}{}_{a} D_{M} V_{A b}\,,
\\
\hPhi_{\bra m a} &= - \brV^{A}{}_{\bra} V^{M}{}_{m} V^{B}{}_{a} D_{M}P_{AB}\,,
\\
\hPhi_{\bra ab} &= \brV^{A}{}_{\bra} V^{B}{}_{a} V^{C}{}_{b} f_{ABC}\,,
\ea
\ee
and $\hbPhi_{\hM \hbm\hbn}$ part: 
\be
\ba{cl}
\hbPhi_{[\brm\brn\brp]} &= \brPhi_{[\brm\brn\brp]} -3  \brV^{M}{}_{\brm} \brV^{N}{}_{\brn} \brV^{P}{}_{\brp} \left(\cA_{[P}{}^{A} \partial_{M} \cA_{N] A} -\frac{1}{3} f^{ABC}\cA_{PA} \cA_{MB} \cA_{NC}\right)\,,
\\
\hbPhi_{[\brm\brn\bra]} &= - \brV^{M}{}_{\brm} \brV^{N}{}_{\brn} \brV^{A}{}_{\bra} (\cF_{MN})_{A}\,,
\\
\hbPhi_{[\brm\bra\brb]} &= \brV^{M}{}_{\brm} \brV^{A}{}_{\bra} D_{M}\brV_{A \brb}\,,
\\
\hbPhi_{{[\bra\brb\brc]}} &= \brV^{A}{}_{\bra}\brV^{B}{}_{\brb}\brV^{C}{}_{\brc} f_{ABC}\,,
\\
\hbPhi^{\hbn}{}_{\hbn \brm} &= \brPhi^{\brn}{}_{\brn \brm} + \brV^{M}{}_{\brm} \cA_{M}{}^{A} f_{A} \,,
\\
\hbPhi^{\hbn}{}_{\hbn \bra} &= \brV^{A}{}_{\bra} f_{A}\,,
\\
\hbPhi_{p \brm\brn} &= \brPhi_{p \brm \brn} - 3 V^{P}{}_{p} \brV^{M}{}_{\brm} \brV^{N}{}_{\brn} \left(\cA_{[P}{}^{A} \partial_{M} \cA_{N] A} -\frac{1}{3} f^{ABC}\cA_{PA} \cA_{MB} \cA_{NC}\right)\,,
\\
\hbPhi_{a \brm \brn} &= - V^{A}{}_{a} \brV^{M}{}_{\brm} \brV^{N}{}_{\brn} (\cF_{MN})_{A}\,,
\\
\hbPhi_{p \brm \bra} &= - V^{M}{}_{p} \brV^{N}{}_{\brm} \brV^{A}{}_{\bra} (\cF_{MN})_{A}\,, 
\\
\hbPhi_{p \bra \brb} &= V^{M}{}_{p} \brV^{A}{}_{\bra} D_{M} \brV_{A \brb}\,,
\\
\hbPhi_{a \brm \bra} &= - V^{A}{}_{a} \brV^{M}{}_{\brm} \brV^{B}{}_{\bra} D_{M}P_{AB}\,,
\\
\hbPhi_{a \bra\brb} &= V^{A}{}_{a} \brV^{B}{}_{\bra} \brV^{C}{}_{\brb} f_{ABC}\,,
\ea
\ee

%%%%%%%%%%%%%%%%%%%%%%%%%%%%%%%%%%%%%%%%%%%%%%%%%%%%%%%%%

\end{document}